\def\roughly#1{\mathrel{\raise.3ex\hbox{$#1$\kern-.75em
\lower1ex\hbox{$\sim$}}}}
\def\lsim{\roughly<}

\documentstyle[prl,aps]{revtex}
\begin{document}
\twocolumn[\hsize\textwidth\columnwidth\hsize\csname@twocolumnfalse%
\endcsname
\vspace{3mm}

\draft
\title{Astrophysical Constraints on Modifying Gravity at Large Distances}

\author{A. Aguirre,${}^a$ C.P. Burgess,${}^{a,b}$ A. Friedland${}^a$
and \underline{D. Nolte${}^a$}}

\address{
${}^a$ School of Natural Sciences, Institute for Advanced Study, 
Princeton, New Jersey 08540, USA\\
${}^b$ Physics Department, McGill University, 3600 University St.,
Montr\'eal, Qu\'ebec, Canada, H3A 2T8.}

\begin{flushright}
\textsl{Dedicated to the memory of Detlef Nolte}
\end{flushright}

\maketitle

\begin{abstract}
{ Recently, several interesting proposals were made modifying the law
of gravity on \emph{large} scales,
within a sensible relativistic formulation. This allows a precise
formulation of the idea that such a modification might account for
galaxy rotation curves, instead of the usual interpretation of these
curves as evidence for dark matter. We here summarize several
observational constraints which any such modification must satisfy,
and which we believe make more challenging any interpretation of
galaxy rotation curves in terms of new gravitational physics.  }
\end{abstract}
\pacs{PACS numbers: \hspace{2in}McGill-01/07}
]

%
%

\section{Introduction}

The observed flatness of galaxy rotation curves are used as important
evidence for the existence of dark matter on galactic scales.  
These observations thus form a keystone of the
structure of inferences that produce the current cosmological
paradigm, which requires the vast majority of the matter in the
universe to consist of unknown, dark forms. Although this paradigm has
received considerable support from recent observations,
this support has been at the
expense of simplicity, since the evidence now requires two forms of
dark matter: both the conventional sort plus the newly-discovered
`dark energy' \cite{CosmologyParadigm,DEevidence}.

The validity of the Newton-Einstein laws of gravity on galactic scales
is an important component of the standard reasoning, and one for which
the direct observational support is comparatively weak. Although much
less work has been done to explore possible modifications of gravity
on large scales than has been invested in dark matter models, over the
years a literature on the subject has
emerged~\cite{modifications}-\cite{mannheim}.  Among the variants is a
reasonably simple modification of Newtonian dynamics which describes
galaxy rotation curves quite well, without requiring the existence of
dark matter \cite{Milgrom1983a,Milgrom1983b,Milgrom1983c,MOND}.

Much of this work has remained outside of the mainstream for various
reasons. Some proposals invent ad-hoc nonrelativistic potentials
without addressing how these potentials might arise within the context
of relativistic field theory. This is not a small omission, because it
is precisely the issue of relativistic consistency which makes
modifying physics at large distances difficult. Although it has long
been known that the laws of gravity are likely to receive
modifications on {\it short} distance scales, general constraints like
unitarity and lorentz invariance in weak fields limit what can be done
on very large scales. These arguments seem to restrict any such
modification to take the form of supplementing the graviton with
fields describing various low-spin degrees of freedom
\cite{GeneralLimits}, leading to scalar-tensor-Maxwell kinds of
theories at long distances.\footnote{These do not include
higher-derivative theories, which generically have runaway modes which
violate the unitarity and stability constraints.}  While some of the
proposed modifications to gravity fall into this category it has
proven difficult to reproduce the galactic rotation curves with these
theories without also running afoul of other constraints
\cite{RelativisticMOND}.

The situation may now be changing, with several recent claims
that self-consistent particle-physics models permit the law of 
gravity to be modified on \emph{large} scales. Several 
recent proposals do so using constructions \cite{BWMods} within the
braneworld picture \cite{ADD,RS}, although none of these
has yet tried to achieve acceptable phenomenology at large
distances. Ref.~\cite{Dvali} aims to describe spiral galaxy
rotation curves assuming the 
existence of extremely light four-dimensional fields trapped onto 
galactic-sized defects. These new developments motivate re-examining 
the astrophysical constraints on modifications to gravity.

It is {\it not} our purpose in this note to present another modification
of gravity at large distance scales. Rather, we put aside general
issues of principle, and indicate what targets proposed models must
hit in order to agree with the extensive astrophysical data on galaxy
dynamics. We shall argue here that the data pose a significant
challenge which must be met by any hypothetical modifications to
gravity, and we believe none of the extant proposals completely meet
this challenge. 

The constraints come 
in several types. First, the systematics of galaxy rotation curves
are observationally found to be described by a universal
shape which depends only on the galaxy luminosity and visible size.
In particular, the huge range in sizes of observed galaxies
precludes theories which modify gravity at a fixed distance for
all galaxies. Second, the galactic gravitational potential for 
spiral galaxies like the Milky Way is measured in 3 dimensions, not 
just in the plane of the galactic disk. Finally, 
additional constraints come from the
Tully-Fisher relation and the x-ray observations of gas in galaxy clusters.
Some of these constraints have been raised elsewhere as objections to
various specific proposed modifications \cite{OtherBounds,GravityTests}. 

We emphasize that our purpose in making this argument is not to argue
that phenomenologically-successful modifications to gravity cannot be
made (although this may in the end prove to be true). Rather, we
intend to provide a convenient summary of the astrophysical
constraints which are relatively unfamiliar to the particle physics
and quantum gravity communities.

\section{Constraints}

We now summarize the most serious astrophysical constraints, whose
successful description we claim to be a necessary (though not
sufficient) condition for the success of
any putative realistic modification to
gravity at large distances.  Since most of the proposed modifications
are motivated to describe galaxy rotation curves, we describe these
first before turning to the other constraints.

\bigskip\noindent{\it 1. Rotation Curve Systematics:} 
Observations on many galaxies indicate rotation velocities, $v_{\rm
rot}$, which for sufficiently large distance, $r$, from the galactic
center, become independent of $r$.  This is to be contrasted with the
Keplerian form, $v^2_{\rm rot} = {GM/r}$, which would be expected for
circular orbits well outside of a concentrated mass distribution, and
which Newton's Law would also approximately predict for galaxies
sufficiently far outside the visible material (in the absence of dark
matter).

A common
way to account for this amongst proposed modifications of gravity is to have 
the gravitational potential be modified from
the Newtonian form $-GM/r$ to a new form: $-GM/r+U(r)$. (Proposals for
$U(r)$ include linear\cite{mannheim1,clow}, logarithmic\cite{kb,Dvali}
or Yukawa--type \cite{Eckhardt} functions of $r$, or have it interpolate between
different asymptotic Newtonian regimes with different values for $G$ 
\cite{complicatedform,drummond}.) The successful comparison of General
Relativity with solar-system and terrestrial measurements \cite{GravityTests}
is then accounted for by ensuring $U(r)$ becomes sufficiently small when
$r < r_0$, where $r_0$ is a new fundamental characteristic distance scale.

For the vast majority of models, the modified force is itself
proportional to $M$ -- {\it i.e.} $U(r) = M \, u(r)$ with $u(r)$ an
$M$-independent function -- because it is obtained by the
superposition of forces exerted by the individual particles of which
the galaxy is made. It follows that any such model crosses over from
Newtonian to exotic force laws for a radius, $r_0$, which is
independent of the properties of the galaxy involved.
\footnote{Ref.~\cite{mannheim} avoids
this conclusion by adding \emph{two} additional force terms, one 
depending on $M$ and
one depending solely on the distance from the galaxy center.}

{\it Any} model for which this crossover radius is universal --
regardless of the details of the force law responsible -- does not
provide a good description of the systematics of galaxy rotation
curves. It does not because it predicts the scale, $r_0$, to be
independent of the luminosity of the galaxy. This conflicts with
observations, for which the transition radius (at which the rotation
curves deviate from the predictions of Newtonian gravity applied to
the visible mass) occurs at radii which are correlated with the
galaxy's optical size,\footnote{We use $R_{\rm opt}=3.2 R_D$, where
$R_D$ is the exponential disk scale length, in accordance with the
definition of Ref. \cite{PersicSalucciStel}} $R_{\rm opt}$. Since
`non-Newtonian' rotation curves have been seen in galaxies that range
over several orders of magnitude in overall luminosity, the observed
transition radius (and optical size) varies over an order of
magnitude, from a few to a few tens of kiloparsecs.

The data indicate that when a galaxy's rotation speed is plotted
against $r/R_{\rm opt}$, the resulting curve is a universal function
which depends only on the total luminosity of the galaxy (see Eq. (14)
and Fig. 10 in \cite{PersicSalucciStel}). In galaxies with low
luminosities, the contribution of dark matter becomes important well
inside the optical radius, while in highest luminosity (very large)
galaxies luminous matter adequately describes the rotation curves to
at least $r/R_{\rm opt}\sim 1$. This behavior is incompatible with modifications of
the gravitational force law \emph{at a fixed distance} from the
source. 

To be completely concrete, consider the case of 
two galaxies, N598 and N801 (with data 
taken from \cite{PersicSalucciStel}). The data show that the rotation curve of the
low-luminosity galaxy, N598, would demand the modification of gravity at
distances less than its optical radius, which is 3.9 kpc. This would cause a
dramatic effect on the rotation curve of the high-luminosity N801, well
inside of its optical radius of 38.4 kpc. No such distortion is observed.
We see that the larger galaxies would exhibit
more -- not less -- dark matter than the small ones at a 
given fraction of their optical radius, were the law of
gravity modified at a fixed distance.  

The existence of modifications to gravity which can successfully
ensure that the modifications become important at distances which
depend on the galaxy involved is demonstrated by the proposal called
MOND \cite{Milgrom1983a,Milgrom1983b,Milgrom1983c}.  MOND's success in describing galaxy rotation curves
relies on its modifying Newtonian {\it dynamics} at a minimum
acceleration, rather than modifying Newton's Force Law above a maximum
distance.  That is, the modification of rotation curves from the
Newtonian prediction does not happen at the same distance from the
center of all galaxies, but occurs at a radius for which the Keplerian
acceleration is of order $a_0 \sim 10^{-10}$ m/s${}^2$
\cite{MOND,Sanders}. Indeed, part of what is remarkable about the
observational success that MOND has in describing observations is the
successful fitting of the rotation curves of many different kinds of
galaxies in terms of a single function describing the crossover from
Newton's law at high acceleration $a$ to its modified version at small
acceleration around $a_0$. Best fits also give very reasonable values
for the mass-to-luminosity ratio, $\Upsilon=M/L$ (and so does not
introduce a new reason to require a new kind of dark matter).

\bigskip\noindent{\it 2. The Tully-Fisher Relation and $M/L$:}
Besides predicting flat rotation curves, any modification of gravity
should also account for the dependence of the rotation curves on galaxy
luminosity. Part of this dependence is summarized by the
Tully-Fisher relation \cite{tully}, which expresses an observational
fact: the luminosity, $L$, of a spiral galaxy is tightly correlated
with the asymptotic limit of its rotation velocity, $v_{\infty} =
\lim_{r \to
\infty} v_{\rm rot}(r)$, according to
\begin{equation}
L \propto v_\infty^\alpha,
\end{equation}
with $\alpha\approx 4$\cite{tfrel}.

While the Tully-Fisher relation (and particularly its small scatter) 
is not completely understood within the dark-matter picture, it can
be roughly derived in simple models where galactic disks collapse with
conserved angular momentum within dark matter halos of a form
predicted by numerical simulations\cite{white}.  But any proposal to
replace dark matter with a modification of gravity should be able to
do better, by directly yielding the T-F relation once the
mass-to-light ratio $\Upsilon$ of the matter constituting the galaxies
is specified. Conversely, to satisfy the Tully-Fisher relation any
modified gravity theory will predict a specific form for $\Upsilon(L)$,
and this prediction can be tested.

For example, all models which modify the position-dependence of the
gravitational potential at large $r$ by supplementing the Newtonian
term with $M \, u(r)$ predict $v_\infty^2/M$ to be purely a function of $r$
for circular orbits. If $u(r)$ for a given
model is designed to approach a constant for large $r$ (to describe
the observed flat rotation curves), then that model predicts $M
\propto v_\infty^2$ \cite{Milgrom1983b}, which is consistent with the Tully-Fisher
relation only if $\Upsilon = M/L \propto L^{-1/2}$. This relation is,
however, strongly constrained because $\Upsilon$ can be estimated both
theoretically and observationally for baryonic matter in galaxies over
a wide range in luminosity.

        Theoretically, $\Upsilon \approx (0.5-4) M_\odot/L_\odot$ in
stellar material is predicted by stellar evolution models that can
reasonably reproduce the colors and spectra of observed spiral
galaxies\cite{TML}.  Similar values can also be directly measured in
the centers of both high- and moderate-luminosity galaxies, where
stars are expected to dominate the mass (and where Newtonian gravity
would presumably hold)\cite{RatSal}. 
More generally, stars are observed to
dominate the mass of known baryonic matter in galaxies -- only in the
smallest dwarf galaxies does the mass in gas approach or (marginally)
exceed that in stars~\cite{GPB}. 

Thus $\Upsilon
\sim (0.5-4) M_\odot/L_\odot$ is expected to hold for
spiral disks of all luminosities. This does not pose a problem for
MOND (for example), which predicts $L\propto v_\infty^4$ and therefore
generally predicts $\Upsilon$ to be independent
of $L$ and to lie within this range \cite{MONDML}.
It is problematic for any theory predicting a
varying $\Upsilon(L)$, since the T-F relation is observed to
hold over four decades
in $L$\cite{GPB}. For example, the prediction $\Upsilon \propto L^{-1/2}$
generic to many models implies that $\Upsilon$ must be 
100 times larger in the least luminous
observed galaxies than in the most luminous ones.  Since the {\em
stellar} populations of dwarf and giant galaxies probably have similar
$\Upsilon$, the T-F relation can be fit only by assuming that $\sim
99\%$ of the mass of the least luminous galaxies 
is in an unknown form ({\it i.e.} neither observed stars or
atomic or molecular gas).  Although it may be possible to
invent forms of baryonic matter in which such large amounts of mass 
might be hidden, this would be rather
unsatisfying in a model constructed to remove the need for dark
matter.

\bigskip\noindent{\it 3. Galaxy Potentials in Three Dimensions:} Since
Newton's law in $d$ space dimensions implies a gravitational potential
which varies as $(1/r)^{d-2}$, it is tempting to try to obtain
logarithmic potentials -- and so constant Keplerian velocities -- 
by trapping the field which mediates this force in some way to
two space dimensions (for which Newton predicts a logarithmic
dependence). At first sight this is a very attractive
possibility, since if a field contributing to long-distance
interactions is trapped on a surface of thickness $\ell \sim$ several
kiloparsecs, then interparticle potentials vary like $1/r$ when $r \ll
\ell$, but cross over (within the plane) to $\log r$ 
dependence for $r \gg \ell$. 
(See ref.~\cite{Dvali} for a recent attempt to construct such
a mechanism.) 

A fundamental prediction of any such mechanism is that galactic
rotation curves should look flat in the directions along the trapping
surface, but the galactic gravitational well should be Newtonian when
examined in directions perpendicular to this surface. Putting aside
questions of how elliptical galaxies (which are not flattened and yet
also appear to have dark matter~\cite{elldm}) might fit into
such a scenario, there are several pieces of evidence concerning our
own galaxy which prospective models must confront.

\medskip\noindent$\bullet$
{\sl Globular Clusters:} Globular clusters are well-defined clusters of
stars which are distributed in an approximately spherically-symmetric way 
about the galactic center. They provide an important constraint on models
of the galaxy because: ($i$) they are not confined to the plane of the spiral arms;
($ii$) they are distributed out to 40 kpc from the galactic center;
and ($iii$) at least 26 of the roughly 150 known globular clusters have 
measurable proper motions relative to extragalactic background objects 
\cite{DaupholeColin,Globulars}, and so have known velocities and positions. 

The constraint on galaxy models arises as follows. Given their known positions
and velocities, it is possible to reconstruct the galactic orbits of many 
globular clusters given a model of the galactic gravitational potential. In
particular, for many models -- including those for which the galactic 
mass does not extend out to 40 kpc or so -- the orbital apocenters ({\it i.e.}
those points most distant from the galactic center) lie 50 to 80 kpc from
the galactic centers. Since most of the time in an
orbit is spent near the orbital apocenter, it is statistically very unlikely
to find less than a handful of clusters further than 40 kpc from the galactic 
center, as is observed \cite{DaupholeColin}.

\medskip\noindent$\bullet$
{\sl The Magellanic Stream:} The Magellanic Stream is a trail of neutral hydrogen 
which extends in a great circle more than 100${}^\circ$ across the sky, starting at
the Magellanic Clouds. This hydrogen is most likely gas which has been stripped 
from the Clouds due to tidal forces, and because the Milky Way's gravitational
potential is much deeper than that of the Clouds, the Stream is distributed along 
the orbit of the Magellanic Clouds themselves. This picture is consistent with
the measured infall velocity of the Stream, which varies nearly linearly as a
function of distance from the Clouds, reaching $-200$ km/s at the end furthest
from the Clouds. The line-of-sight velocity of the center of mass of the 
Magellanic Clouds themselves is 61 km/s, indicating that the Clouds are probably
near the pericenter -- closest approach -- of their orbit, which must be quite
eccentric. 

The constraint on the Galactic potential is obtained by requiring the
existence of an orbit with a pericenter of roughly $q = 50$ kpc and on
which radial velocities as large as 200 km/s are obtained
\cite{FichTremaine}.  This puts strong constraints on models with all
of the Milky Way's mass localised where the luminous matter is, and in
which gravity is Newtonian out of the galactic plane, since for these
the Stream motion should be approximately Keplerian. Since the maximum
radial velocity of a bound Keplerian orbit with pericenter $q$ is
$v_{\rm max} = (GM/2q)^{1\over 2}$, for $q = 50$ kpc, the observed 200
km/s infall of Stream material implies a galactic mass $M > 9 \times
10^{11}$ solar masses. This is an order of magnitude larger than the
mass obtained by counting the luminous matter in the disk, but is
consistent with the amount of dark matter required to account for
rotation curves.

\medskip\noindent$\bullet$
{\sl Local Group Galaxies:} Constraints similar in spirit to, but weaker
than, the one just described can be obtained from the dwarf galaxies which
make up much of our local galactic neighborhood. A number of arguments indicate
that these are bound to our galaxy \cite{FichTremaine} and their observed
distances and radial velocities constrain the shape and depth of
the galaxy's gravitational well. In particular, the large speed of some of these
dwarfs (such as Leo I, Pal 14 and Eridanus) can only be bound by the
presence of much more mass than is visible if the out-of-plane 
gravitational field is Newtonian.

\bigskip\noindent{\it 4. 
The Dynamics and Structure of Clusters of Galaxies:}
        As first notice by Zwicky~\cite{zwicky}, the visible mass in
clusters of galaxies is entirely too small to bind them, given the
observed velocity dispersion (several hundred km/s) in their component
galaxies; this was the first strong evidence for dark matter (or
modified gravity) in cosmology, and is yet stronger today.  X-ray
measurements of clusters have revealed that their observable baryonic
masses are dominated by hot gas of temperature $\sim 1-10\,$keV, which
also implies (using standard gravity) a binding mass well above that
observed in galaxies and the hot gas itself.  

It is instructive to derive this inconsistency explicitly, in
order to see the logic of the constraints posed on dark matter and on
modifications of gravity. Consider, then, the
equation of hydrostatic equilibrium in a spherical system (which
describes intracluster gas well):
\begin{equation}
{1\over\rho}{dP\over dr}=-a(r),
\end{equation}
where $\rho$ and $P$ are the density and pressure of the gas, and $a$
is the inward gravitational acceleration at radius $r$.  
Using the ideal gas equation of state, this
can be re-written in terms of the gas temperature $T$ and mean
molecular weight, $\mu \sim 0.6$ (in units of the proton mass, $m_p$):
\begin{equation}
{d\log \rho\over d\log r}+{d\log T\over d\log r}=
        -{r\over T}\left({\mu m_p\over  k}\right)a(r),
\label{eq-hyel}
\end{equation}
where $k$ is Boltzmann's constant. Clusters are observed to have
temperature profiles, $T(r)$, which are roughly constant outside of
their cores. The density profile of the observed gas at large radii
roughly follows a power law, $\rho \propto r^\alpha$, with index $-2
\lsim \alpha \lsim -1.5$\cite{carameli}. Using these, the fact that
the gas dominates the cluster's baryonic mass, and Newton's expression
for $a(r)$ in the absence of dark matter gives
\begin{equation}
kT \approx (1.3-1.8){\rm keV}\left({M_r\over 10^{14}\,M_\odot}\right)
\left({1\,{\rm Mpc}\over r}\right)
\end{equation}
for the baryonic mass within the central $1\,$Mpc of a typical rich
cluster, where $M_r$ is the mass enclosed within $r$.  The disparity
between this estimate and the corresponding observed temperature
$\approx 10\,$keV~\cite{mbrel} indicates the need for dark matter or
modified gravity.

Similar reasoning provides a strong constraint on any modifications of
gravity, for which the only matter present is visible and $a(r)$ is
calculable.  Most generally, Eq.~(\ref{eq-hyel}) implies an
approximate relation between the total mass, outer radius, and average
temperature of the cluster.  For a given gravity theory, this relation
can be checked by comparing to observed masses, temperatures, and
sizes of observed clusters, as we have done for Newtonian gravity
above.  A useful compilation of applicable cluster data is given
in~\cite{mbrel}.  MOND can apparently survive this
comparison~\cite{aguirre}, as can the standard CDM cosmology with
$\sim 5-10$ times as much dark matter as baryonic matter, whereas (for
example), the construction of~\cite{Dvali} would not account for the
high cluster temperatures (and hence would still require the standard
amount of dark matter) because most cluster gas would {\em not} lie
within the domain walls and would hence feel only the usual Newtonian
gravitation of the visible matter.

        A stronger constraint can be deduced using observed density,
mass and temperature \emph{profiles} of individual clusters. If
gravity depends only on the visible mass, the observed density profile
$\rho(r)$ and enclosed mass profile $M_r$ of a given cluster can be
used to directly predict its temperature profile $T(r)$ in a given
gravity theory using Eq.~(\ref{eq-hyel}), and the profile compared to
the measured one.  Roughly speaking, the combination $r a(r)$, which
is proportional to $T(r)$ (since $d\ln \rho/d\ln r$ and $T$ are 
observed to be approximately constant) should be nearly independent of
$r$, as $T(r)$ is observed to be.  This technique is illustrated in
detail in~\cite{aguirre}, where it is shown that MOND predicts rising
radial temperature profiles that disagree badly with the observations
(which show that clusters are roughly isothermal).  

Other modifications of gravity may encounter similar difficulties.  
Consider, for example, the hydrostatic equilibrium of cluster gas 
when large-$r$ gravity is dominated by a linear 
potential~\cite{mannheim1,mannheim}.  
Since linear potentials are sensitive to the mass distribution at large
radius, we will assume the cluster gas to not extend beyond an outer
radius, $R$.\footnote{Given the extreme sensitivity of the interaction
energy to matter at arbitrarily large radii, it is unclear how to compute 
the characteristics of a given physical system with a linear potential
without neglecting this contribution.} For simplicity of analysis we
bracket the real mass density for $r<R$ between two extremes: it is taken to be 
concentrated at radius $r=0$, or to be uniform within $R$. Under
these assumptions we find that for $r \ll R$, $r a(r)$ respectively grows 
either linearly or quadratically with $r$. Provided only that
a more realistic treatment lies between these two extremes, we
see that $T(r)$ is predicted to increase linearly to quadratically 
with $r$. Such a prediction would be ruled out by the observations, 
although a more detailed study would be required to make
this conclusion fully quantitative.

Finally, we note that cluster mass estimates using galactic dynamics
or the X-ray gas are consistent (to within a factor of two) with
masses determined by weak gravitational lensing of background galaxies
around the clusters~\cite{clens}. Thus any alternative gravity theory
must also predict gravitational lensing in a way that allows this.
This constraint was used, for example, by~\cite{kb} to rule out their
model, and provides difficulties for MOND in cluster
cores~\cite{MOND}.

\bigskip It would be wonderful to discover new gravitational physics
from astrophysical observations, and it is intriguing that current
braneworld models seem to imply that long-distance gravitational
physics can be much richer than had been previously thought. In order
for any such physics to provide a viable alternative to dark matter it
must satisfy the above constraints, and a theory that successfully
does so would be of sufficient interest to merit a more detailed investigation
(including detailed rotation-curve fitting, lensing calculations,
predictions concerning the formation of structure, cosmological
predictions, etc.).  We hope our summary of these constraints will
further stimulate thinking about the relation between astrophysical
puzzles and current ideas in high-energy physics.

We are grateful to Gregory Gabadadze for his valuable comments. This research
was partially funded by grants from F.C.A.R (Qu\'ebec),
N.S.E.R.C. (Canada), as well as the Ambrose Monnell and the W.M. Keck
Foundations.

\end{document}